\documentclass[sigconf]{acmart}

\usepackage[brazil]{babel}   
\usepackage[utf8]{inputenc}
\usepackage{tikz-cd}
\usepackage{lipsum}
\usepackage{wasysym}
\usepackage{multirow}

\usepackage{hyperref}
\usepackage{array}                  
\newcolumntype{L}[1]{>{\raggedright\let\newline\\\arraybackslash\hspace{0pt}}m{#1}}
\newcolumntype{C}[1]{>{\centering\let\newline\\\arraybackslash\hspace{0pt}}m{#1}}
\newcolumntype{R}[1]{>{\raggedleft\let\newline\\\arraybackslash\hspace{0pt}}m{#1}}
\newcolumntype{J}[1]{>{\let\newline\\\arraybackslash\hspace{0pt}}m{#1}}

\AtBeginDocument{%
  \providecommand\BibTeX{{%
    \normalfont B\kern-0.5em{\scshape i\kern-0.25em b}\kern-0.8em\TeX}}}

\begin{document}

\title{Parthenos: A Source Code Injection Approach for Software Transformation}

\author{Gabriel Lopes Nunes}
\affiliation{%
  \institution{Universidade do Vale do Rio dos Sinos}
  \city{São Leopoldo}
  \state{Rio Grande do Sul}
}
\email{gabriellopes.nunes@hotmail.com }

\author{Kleinner Farias}
\affiliation{%
  \institution{Universidade do Vale do Rio dos Sinos}
  \city{São Leopoldo}
  \state{Rio Grande do Sul}
}
\email{kleinnerfarias@unisinos.br}

\author{Lucas Silveira K{\"{u}}pssinski}
\affiliation{%
  \institution{Universidade do Vale do Rio dos Sinos}
  \city{São Leopoldo}
  \state{Rio Grande do Sul}
}
\email{lkupssinsku@edu.unisinos.br}

\begin{abstract}
Maintaining legacy enterprise information systems is a known necessity in companies. To date, it remains an expensive and time-consuming process, requiring high effort and cost to get small changes implemented. MITRAS seeks to reduce the maintenance cost by providing an automatic maintenance system model based on graph transformations. This article presents Parthenos, a different approach to MITRAS, whose goal is to guarantee the correctness of introduced modifications at a syntax and type semantics level of the source code. Along with that, it proposes an extensible architecture, which allows the most varied types of systems to carry out software maintenance. Parthenos was evaluated through functional tests to evaluate its effectiveness, using measures of precision, recall, and f-measure. 
\end{abstract}



\maketitle

\section{Introduction}
\label{sec:introduction}


Operating and maintaining systems is costly for both customers and companies behind such processes~\citep{espindola2004}. The ever-present changes in software requirements cause architectural instability~\cite{farias2014effects, bischoff2019integration}, require high maintenance effort~\cite{farias2013analyzing, farias2012evaluating} and impair the understanding of the source code~\cite{gonccales2019measuring, gonccales2021measuring, vieira2020usage}. Even though it represents the largest portion of the cost from software projects, maintenance receives a lot of attention from companies~\cite{d2020effects}, as they usually have profitable legacy software systems that need to continue to function properly over time. Often, customers demand for small changes, which require companies to work with them in parallel with other newer projects~\citep{kupssinski2019}. This way, customers may not have their needs met in time, or they will need to spend a lot more money for the desired changes to be implemented faster~\citep{espindola2004}. To mitigate these types of problems, K{\"{u}}pssinski~\citep{kupssinski2019} developed MITRAS.


MITRAS introduces a software model that uses formalities for manipulating and transforming graphs to give systems the ability to enable end-users, who do not have programming knowledge, to perform adaptive and perfective maintenance on them~\citep{kupssinski2019}. MITRAS is carried out along three main steps: model extraction, transformation and source code injection. It also has a module for end-user communication that uses natural language processing (NLP) to understand which maintenance the user wants to perform. It then applies the necessary transformations and injections~\citep{kupssinski2019}. Currently, MITRAS proposes a limited injection process, specifically for Java, JSP and Hibernate. Another handicap is its disregard for data typing, making all types be taken as \textit{string} in both transformation and injection steps. Moreover, it lacks metadata in the model, which makes it difficult to trace the source code from the graph. 


For these matters, this article proposes Parthenos, an approach to guarantee the correctness of introduced modifications at a syntax and type semantics level of the source code. That is, it allows for ensuring the syntactic and type-semantic correctness of systems throughout the transformation process and keeping the synchronization between the involved models. Furthermore, Parthenos also provides a level of extensibility in its architecture that allows systems using different languages, frameworks, libraries and tools to take advantage of such types of maintenance.


The rest of the article is structured as follows: Section~\ref{sec:background} presents the theoretical foundations that are considered necessary to understand the subsequent sections. Section~\ref{sec:related_works} presents and compares related works. Section~\ref{sec:modelo} introduces Parthenos, describing the model and its proposed architecture. Section~\ref{sec:evaluation} defines the evaluation methods and scenarios, explains the metrics used, and presents the obtained results. Finally, Section~\ref{sec:conclusion} describes aspects of the work and presents suggestions for future work.

\section{Background}
\label{sec:background}

This section covers the theoretical concepts used during the construction and development of Parthenos. 

\subsection{Graph}
\label{subsec:graph}


Graphs are very visual mathematical representations. They are used to represent models, which have points and lines that connect some of these points two by two~\citep{bondy1976}. This representation is so generic that it can be used to model a variety of types of real-world situations, where one needs to portray relations between elements. Thus, it is evident that graphs are great tools for modeling binary relations~\citep{knauer2011}.


A graph can be defined as a triple $G = (V,E,p)$, $V \neq \emptyset$, where $V$ is a finite set of vertices, $E$ is a set of edges and $p : E \rightarrow V^{2}$ is a mapping that relates some edge $e \in E$ with a pair of vertices $(u, v) \in V^{2}$, and represents the connection of $u$ and $v$ by a line~\citep{knauer2011}. It is possible to denote $G$ as $(V_{G}, E_{G},p_{G})$ or $(V(G), E(G), p_{G})$, where $p_{G}(e)$ = $uv$, to identify to which graph each element of the triple belongs~\citep{bondy1976}.


To convey other types of semantic information, it is possible to modify the structure of a graph. It can be imposed, for example, that the edges of a graph are arrows to represent unidirectional relations; these graphs are then called digraphs or directed graphs. Or one can associate labels or weights to the vertices and edges of a graph, which are called labeled graphs~\citep{gersting2006}. MITRAS model uses a directed and labeled graph, which Parthenos shares, so that it is possible to identify hierarchies of classes or database diagrams~\citep{kupssinski2019}.

\subsection{Graph grammars}
\label{subsec:graph_grammars}

Graph grammars are a generalization of the theory of string-based formal languages and the theory of tree-based rewriting systems. In the way these grammars are defined, it is easy to represent local transformations in graphs in a mathematically precise way, which allows inferring properties and proving certain aspects of them more easily~\citep{Rozenberg1997}.

Grammars are generically composed of productions, which are the rules to be followed to carry out the transformation of a graph. These productions are seen as a triple $(M, D, E)$, where $M$ is the mother graph, $D$ the daughter graph and $E$ is some reinsertion mechanism, such that the production can be applied to some graph $G$, whenever an occurrence of $M$ is identified or matched in $G$. As a result, by applying the production, $M$ is removed from $G$, obtaining a graph $G^{-}$, in which $D$ is reinserted using the mechanism $E$. There are several types of reinsertion mechanisms that can be considered when applying a production, two of which are the most important: the adhesive and the connective. In the first approach, parts of the graph $D$ are mapped to $G^{-}$, while in the second, new edges are created to connect $D$ to $G^{-}$. Based on these mechanisms, two important approaches for rewriting graphs are defined: algebraic, based on Category Theory, and algorithmic, based on Set Theory~\citep{Rozenberg1997}.

Parthenos makes use of the algebraic approach, employing, more specifically, the single pushout technique to apply transformations. In this technique, a transformation (matching, deletion, and insertion of vertices and edges) can be done in a single step as a construct of two morphisms $p$ and $m$, where $p$ is a partial morphism and $m$ is a total morphism of graphs.




\subsection{MITRAS}
\label{subsec:mitras}



\begin{figure}[!ht]
\centering
\includegraphics[scale=0.67]{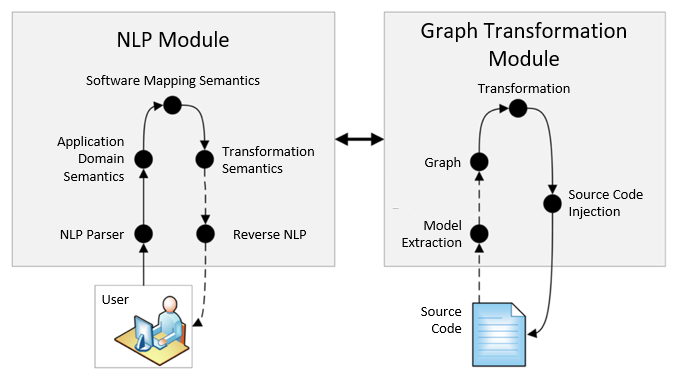}
\caption{Overview of the MITRAS model~\citep{kupssinski2019}.}
\label{fig:MITRAS}
\end{figure}

MITRAS is a model for automated software maintenance based on natural language processing and graph transformations that aims to provide the end-user of a given application the power to perform maintenance without prior knowledge in programming~\citep{kupssinski2019}. Figure~\ref{fig:MITRAS} encompasses three steps and a module for end-user interaction.

\textbf{Step 1: Source code scanning.} MITRAS scans the application\rq s repository to find important files for its representation. Files of interest are those containing classes, settings, persistence information, and UI-related. The identified dependencies, associations, among other, are all mapped in an abstract graph model, which depicts the application at a sufficiently high level, so that it is still possible to track the source code from it, but avoids exposing complexities that are unnecessary for the execution of the process as a whole~\citep{kupssinski2019}.

\textbf{Step 2: Transformations.} The abstract graph is ready to receive transformations. Four types of transformation are proposed: adding a field, hiding a field, unhiding a field and repositioning a panel. Following the algebraic approach to graph transformations, before applying a transformation, a subgraph $D$ will be identified in the original graph through a morphism. The transformation is applied over the identified subgraph following the rule defined in a production $p$ that takes $D$ to some graph $D^{'}$ that can have edges and vertices inserted or removed. The graph $D^{'}$ is then reconnected to the original graph following the chosen reinsertion policy, which is that of creating new edges.

\textbf{Step 3: Injection.} all transformations performed in the abstract graph are translated into code and inserted in the application\rq s source files. This step is called injection, as MITRAS model would insert code snippets, or change the application's code at specific points in the source file. This way, they would not all be rewritten, and the synchronization between the abstract model and the source code of the application would be done with a good performance.

\section{Related Works}
\label{sec:related_works}


The search for related works was done through digital repositories such as Google Scholar. Some of the works presented were found by searching for ``source code transformation'', ``automatic code transformation'' and ``automatic code refactoring'' search strings, while others were obtained by convenience.

\begin{table*}[!ht]
    \footnotesize
	\caption{Comparison of the related work.}
	\label{sample-table}
	\centering
	\begin{tabular}{ccccccc}
	
	\toprule
        \multicolumn{1}{c}{\textbf{Selected Studies}} & 
        \multicolumn{1}{c}{\textbf{Abstraction Level}} & 
        \multicolumn{1}{c}{\textbf{Abstract Model}} & 
        \multicolumn{1}{c}{\textbf{Transformations}} & 
        \multicolumn{1}{c}{\textbf{Type Concerned}} & 
        \multicolumn{1}{c}{\textbf{Semantic Correctness}} & 
        \multicolumn{1}{c}{\textbf{Assessment Type}}\\
    \midrule 

\multicolumn{1}{l}{Parthenos}               & High                 & Yes               & M2T                                      & Yes                & Yes                 & Functional test                  \\ 
\multicolumn{1}{l}{Kupssinsk{\"u} \citep{kupssinski2019}}                              & High                 & Yes               & T2M, M2M, & No                 & No                  & Use case,    \\ 

& & & M2T & & & performance test       \\ 

\multicolumn{1}{l}{Chagas et al. 2019 \cite{chagas2019}}   & High                 & Yes               & M2M                                        & No                 & No                  & Questionnaire, \\ 

& & & & & & performance test  \\

\multicolumn{1}{l}{Antal et al. \cite{antal2016}}                            & Low                  & No                & T2T                                        & No                 & No                  & Functional and \\
& & & & & & performance tests \\

\multicolumn{1}{l}{Zafeiris et al. \cite{Zafeiris2017}} & Middle               & No                & T2T                                        & No                 & Yes                 & Functional test                  \\
\multicolumn{1}{l}{Oliveira et al. \cite{oliveira2018}}                         & Middle               & Yes               & M2T, T2T                                     & Yes                & Yes                 & Case study                       \\

\multicolumn{1}{l}{Gil and Orr{\`u} \cite{gil2017}}       & Middle               & No                & T2T                                        & No                 & Yes                 & Case study                       \\ 

\bottomrule
\end{tabular}

\end{table*}

\subsection{Analysis of related works}
\label{subsec:analysis_of_RW}


\textbf{Kupssinsk{\"u} \citep{kupssinski2019}.} This study discusses MITRAS model and shows the feasibility of using the model, as well as how it performs in realistic scenarios. Despite covering all the stages on which the model is based, from extracting a repository\rq s model to injecting the source code, the focus of the work is on demonstrating its usability. However, there is little discussion on the injection step. The necessary tools and the challenges of injecting the changes introduced by the transformations in the source code are not addressed. Besides, the prototype implements this step in a very limited way.

\textbf{Chagas et al. 2019 \cite{chagas2019}.} The article deals with an external MITRAS module that is responsible for communicating with the end-user through NLP. It uses CoreNLP to mark part-of-speech of the requested sentences and uses an ontology modeled through OWL language that represents the specific concepts of the analyzed scenario, such as entities and UI panels to link words and their interrelations to users\rq intentions. After that, it invokes MITRAS' transformation module providing the necessary parameters. It does not model typing relations such as, for example, understanding that an ``age'' field has a numeric type. Such relations are essential for the injection step to be semantically correct.


\textbf{Antal et al. \cite{antal2016}.} This work presents a framework for transforming C++11 code to C++03 code. The idea of the project is to provide a way to enable developers to use a more recent version of the language, leveraging the sugars it has, without affecting interoperability with legacy systems. The solution, which is open source, has a list of several transformations that can be applied to certain types of constructs in C++11. The system preprocesses the source files to find the adequate transformations, transforms the files, and maintains meta-information to track versions of the transformed source code. 

\textbf{Zafeiris et al. \cite{Zafeiris2017}.} This work develops an automatic refactoring of Java classes from Call Super design pattern to Template Method. The first pattern is very common in object orientation. When one wants to extend the behavior of a concrete method of the parent class, a call to it is made using the \textit{super} keyword from a method of the same name contained in the child class. The second pattern belongs to the set of behavioral design patterns. It introduces more well-defined extension points, where greater control over the behavior of the method to be extended can be obtained. 

\textbf{Oliveira et al. \cite{oliveira2018}.} This article proposes BRCode, a solution that uses an interpretive approach in MDE (model-driven engineering) to assist in application development. BRCode interprets the models and metadata created by the developer to generate an application that already has several built-in functions, such as an authorization and authentication layer and an internationalization layer. It resembles MITRAS for interpreting and dealing with models to assist in the automatic development of applications. However, BRCode works with a medium level of abstraction, since it deals with metadata and models that are very close to the application they represent.

\textbf{Gil and Orr{\`u} \cite{gil2017}.} This work presents Spartanizer, a plugin for Eclipse that applies automatic refactorings in the source code, to adapt it to Spartan programming. Such programming style values minimalism and lean writing in codes. Thus, it prefers short to long variable names, ternary operators to conditional blocks, etc. Spartanizer parses the source code and identifies refactoring suggestions. Spartanizer can refactor the source code successively through transformations. These transformations are previously identified and maintain the semantic correctness of the code in most cases. 

\subsection{Comparative analysis of works}
\label{subsec:compativo}

\textbf{Comparison criteria.} To contrast the related work some comparison criteria were defined as follows: (1) \textit{Abstraction level:} It informs the level of abstraction the solution presents to apply the transformations, being the source code the lowest level; (2) \textit{Abstract model:} It indicates if the solution uses some type of abstract model where the transformations are applied; (3) \textit{Transformations:} It lists the types of transformation that the solution presents. Each work can present types among text-to-text (T2T), text-to-model (T2M), model-to-model (M2M), model-to-text (M2T), according to transformation classifications proposed in~\citep{kahani2011}; (4) \textit{Type concerned:} It indicates whether the solution is concerned with dealing with types in strongly typed languages; (5) \textit{Semantic correctness:} It indicates whether the solution tries to maintain semantic correctness between transformations; and (6) \textit{Assessment type:} It describes how the solution was validated.

\textbf{Research Opportunity.} Table~\ref{sample-table} presents the comparison of the related work considering these criteria. Some research opportunities were derived from this comparison from which Parthenos was designed. This way, Parthenos proposes a solution for the source code injection step that covers these gaps, aiming to improve the use of MITRAS model by: (1) providing a mechanism to persist the changes made to the models after a transformation and ensuring their synchronization; (2) performing the injection and guaranteeing the type semantics and syntactic correctness of the source code; and (3) providing mechanisms that allow different types of systems to use the model, through the development of extensions that support them.

\section{The Proposed Model}
\label{sec:modelo}


This section presents the Parthenos, a 3-step approach to guarantee the correctness of introduced modifications at a syntax and type semantics level  of source code. 

\subsection{Overview of the proposed process}
\label{subsec:modelo} 


Parthenos aims to solve the existing problems in the source code injection step of MITRAS model. However, to inject the source code, maintaining a correct type semantics and syntax, it is necessary to make several improvements in the way~\citep{kupssinski2019} solution implements the extraction, transformation, and injection steps. This is because there is no concern with the types in them: the abstract graph model does not generate relations with typings of fields or methods, the transformer does not receive the type of new fields or methods to add or change and, as a consequence, the injector assumes all of these as being of type \textit{string}. Furthermore, it is impossible for systems written in languages other than Java to be maintained by MITRAS. Thus, it was decided to reimplement the main steps it presents: extracting the model, transforming, and injecting the source code~\citep{kupssinski2019}. Thus, Parthenos makes use of the model described by MITRAS, but provides a new implementation. 

\begin{figure}[!h]
\centering
\includegraphics[scale=0.32]{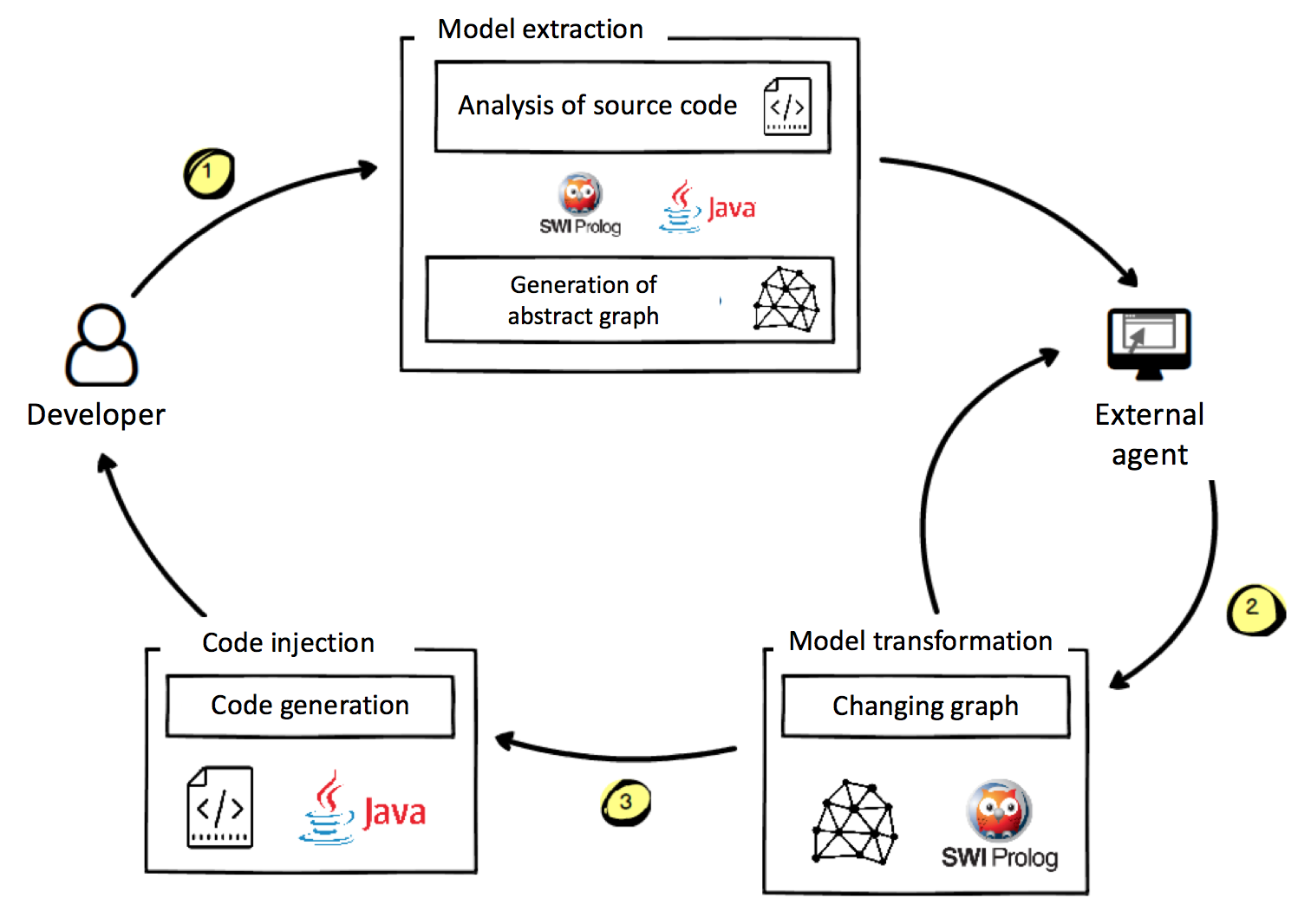}
\caption{Overview of the proposed model.}
\label{fig:process}
\end{figure}

Figure~\ref{fig:process} presents an overview of the proposed approach. A description of each is given as follows:

\begin{itemize}


\item \textbf{Step 1: Model extraction.} In this step, a developer, or a user with development knowledge, extracts the abstract graph model from the system's repository. The repository of this system is scanned by Parthenos and all source files that are important for its representation, such as class files, UI-related files, and configurations, are collected, analyzed and a graph containing the necessary information for the representation of this system is generated. The graph is made available so that other applications can use it to apply transformations or obtain information and metadata from the system represented by it.


\item \textbf{Step 2: Model transformation.} This step is triggered when an external agent needs to transform the graph produced in the previous step. The agent communicates to Parthenos the transformation it wants to perform and points to the graph, where the transformation will be applied. Parthenos changes the graph and makes it available for future transformations. After that, the source code injection step is initiated so that the source code is synchronized with the abstract graph model. This step is performed several times, whenever it is necessary to introduce a system maintenance.


\item \textbf{Step 3: Source code injection.} In this step, the changes made in the graph will be synchronized with the application's source file repository. It ensures that both models are correct and in synchrony, so that it is possible to improve the application through successive transformations. This step is invoked by Parthenos transparently in Step 2. Therefore, whenever Step 2 is executed, Step 3 is also triggered, thus both the graph and the source code are changed. Parthenos informs which transformation needs to be injected and points to the repository where the source files are located. The respective code is generated and saved in the repository.

\end{itemize}

Parthenos focuses on presenting a way to inject into the source code the changes that are introduced when some transformation is applied. To make this possible, it performs a series of steps along the whole process, that are necessary to enable the injection to be made. The steps are as follows: Find the changes introduced by a transformation; Identify the files that need to be injected; Identify the type of injection to be done; Obtain, from the graph, the necessary parameters to perform the injection; Find the injection point within each file, considering its grammar; Perform the injection; and Persist the applied transformation in the abstract graph model.


\subsection{Proposed Architecture}
\label{subsec:architecture}

\begin{figure}[h!]
\centering
\includegraphics[scale=0.35]{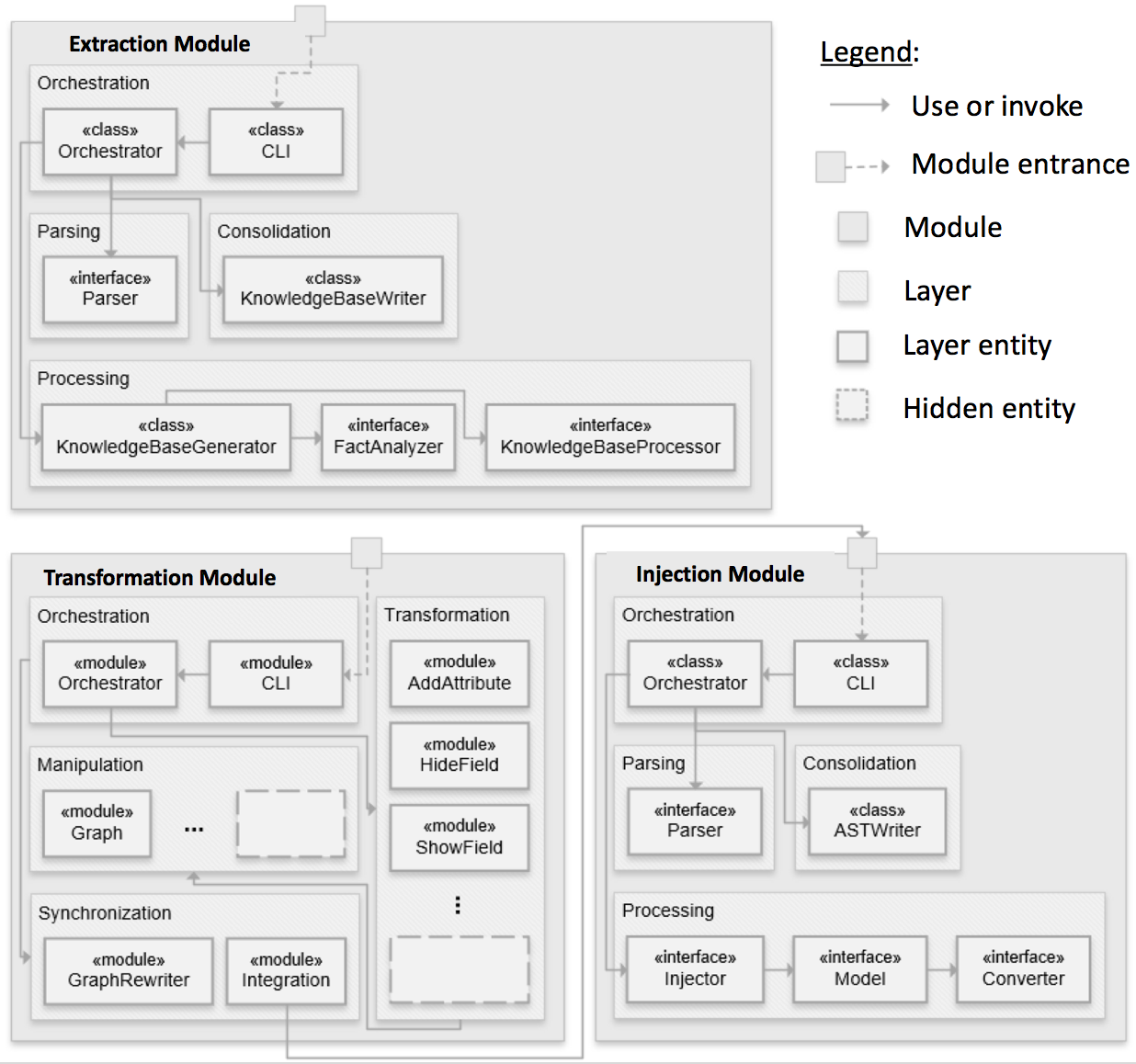}
\caption{The proposed architecture.}
\label{fig:architecture}
\end{figure}

The proposed architecture encompasses the three steps listed in Section~\ref{subsec:modelo}. As seen in Figure~\ref{fig:architecture}, it has three distinct modules, which are responsible for carrying out a specific step. The modules are made up of layers, which aim to facilitate not only the implementation, since they separate the different flows of each step properly, but also the extension of the modules. This architecture regards the core of Parthenos. It does not include built-in extensions, therefore, it will not perform any of the steps out-of-the-box. Such extensions need to implement the interfaces provided in the core, to add proper functionality for specific languages and tools.

\begin{figure}[h!]
\centering
\includegraphics[scale=0.14]{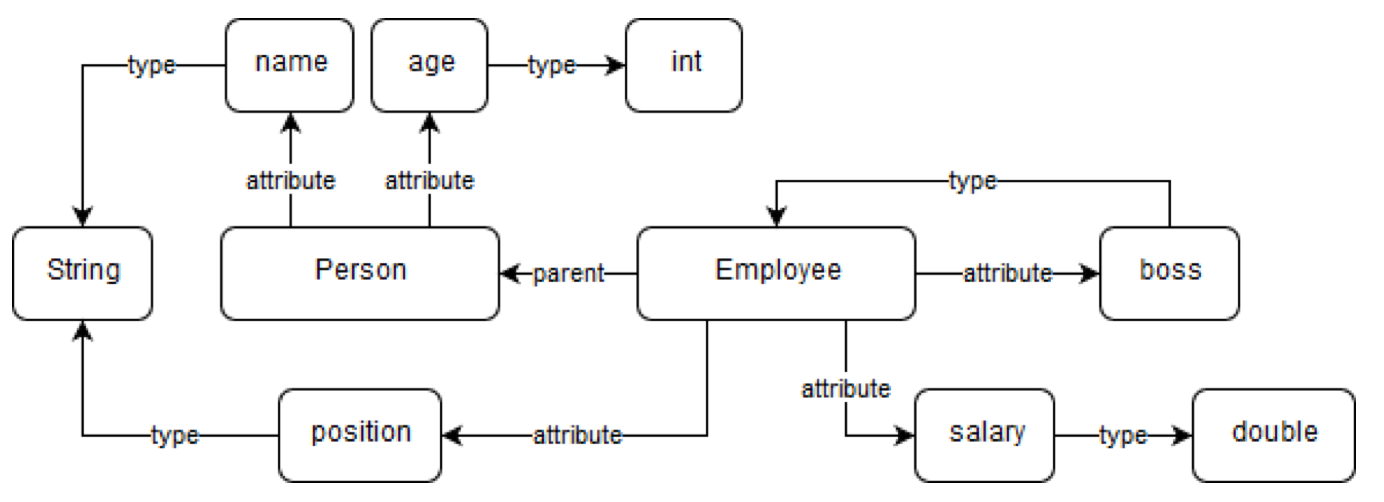}
\caption{Example of a graph obtained from the extraction module.}
\label{fig:example1}
\end{figure}

\textbf{Extraction module.} This module is responsible for generating the abstract graph model. It is written in Java language and is invoked via the command line. The input of the extraction module is a repository of source files, the languages that should be considered for generating the graph, and a destination file path where the graph will be persisted. The extraction module has four layers: orchestration, parsing, processing, and consolidation. The first layer is responsible for commanding the application, interpreting the command line, and invoking the other layers. The second layer deals with parsing the source file into an abstract syntax tree (AST). It does this through the Parser interface. The third layer works with AST processing, to generate the abstract graph model. For this, it uses a \textit{KnowledgeBaseGenerator} class, which generates, with the aid of two interfaces, the knowledge base containing the graph, whose example can be seen in Figure~\ref{fig:example1}. These interfaces are: 1) \textit{FactAnalyzer}: analyzes the AST and generates facts in graph form, thus constituting the knowledge base; 2) \textit{KnowledgeBaseProcessor}: processes the knowledge base, adding metadata and other information. The fourth layer has the purpose of translating the facts in Prolog language and writing the output file, which is done through the \textit{KnowledgeBaseWriter} interface. Through these provided interfaces, it becomes possible to extend the extractor to support various types of languages that may exist in the repository. The output of the extraction module is a file written in Prolog that contains the knowledge base.

\begin{figure}[h!]
\centering
\includegraphics[scale=0.15]{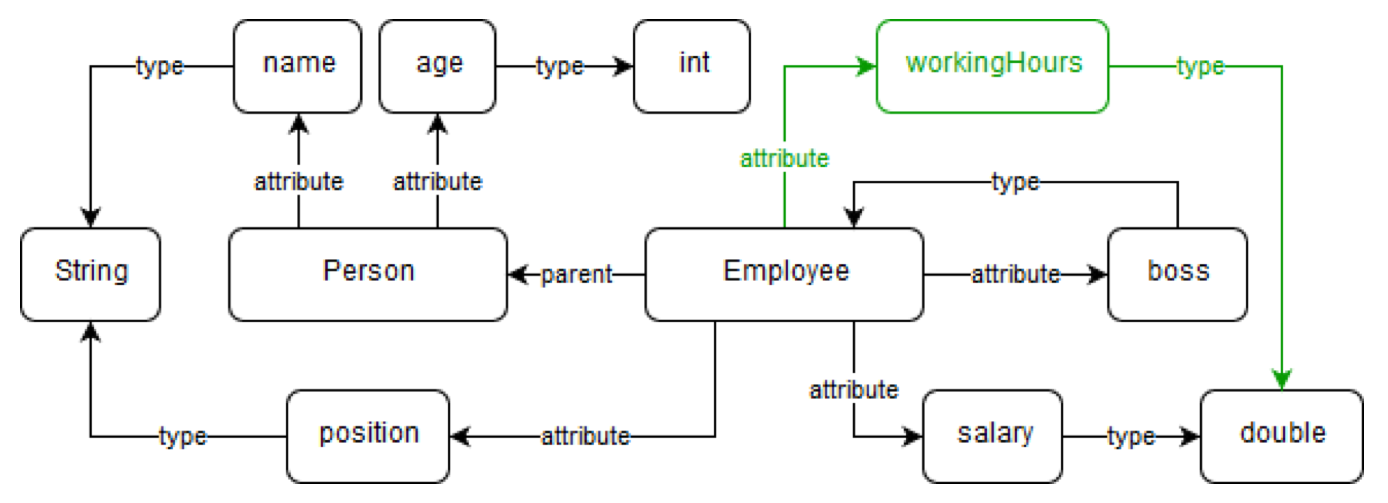}
\caption{Example of a graph obtained from the extraction module.}
\label{fig:example2}
\end{figure}

\textbf{Transformation module.} The transformation module is responsible for transforming the graph generated by the previous module. It is written in Prolog and is also triggered via the command line. The input of this module is the file that represents the abstract graph model (knowledge base), the type of transformation to be performed, and the parameters for this transformation. The module is divided into four layers: orchestration, transformation, manipulation, and synchronization. The first layer, as in the previous module, interprets the command line and orchestrates the flow of the transformation application. The second layer holds the implemented transformations. Each transformation is responsible for matching the graph and, subsequently, applying the necessary changes. The third layer provides the first one with mechanisms to manipulate the abstract graph model, in addition to the algorithms that allow extracting information from it. The fourth layer is responsible for the synchronization of the graph and the source code. It provides, alongside with graph synchronization mechanisms, the integration with the injection module, identifying the parameters to perform the injection and invoking it with the necessary arguments. The output of this module is the modified abstract graph model as seen in Figure~\ref{fig:example2}.

\textbf{Injection module.} The injection module tries to inject the transformations in the source code, thus synchronizing the concrete part of the system. It is written in Java and, although it can be invoked via the command line, it is triggered transparently by the transformation module. The input for this module is the source files to be injected, the languages in which they are written, the type of injection, and a JSON model that contains the parameters for injection. The injection module is divided into four layers just like the extraction module: orchestration, parsing, processing, and consolidation. Of these layers, only the last two perform different functions in this module. The processing layer is responsible for interpreting the JSON model and performing the injection in the AST provided by the parsing layer. Three interfaces are used for this: 1) \textit{Model}: interprets and supplies the parameters contained in the JSON model; 2) \textit{Converter}: its function is to convert the parameters contained in the Model to values understood by the Injector interface; 3) \textit{Injector}: it makes use of the Model to inject the transformation into the AST. The consolidation layer has the role of writing the changes made in AST to the original source file, using the ASTWriter interface. The output of this module is the changed source files.

\subsection{Implementation Aspects}
\label{subsec:implementation_aspects}

The architecture was implemented as the core of Parthenos, providing mechanisms and APIs that facilitate the development of extensions for it. The extensions add functional behavior to Parthenos, since, alone, it cannot generate a graph, transform or inject the source code of a specific language or tool. This way, two sets of extensions were developed for Parthenos, each one providing extensions for the extraction, transformation, and injection modules of the model separately.


The first set of extensions provides support for the Java language. It enables the abstract graph model to contain the representations of classes and their relations attributes, methods, in addition to their metadata. It also adds two transformations: create a class and add an attribute (and its accessor methods get and set). In this set of extensions, Java was used to extend the extraction and injection modules, while Prolog was used for the transformation module.

\begin{figure}[h!]
\centering
\includegraphics[scale=0.15]{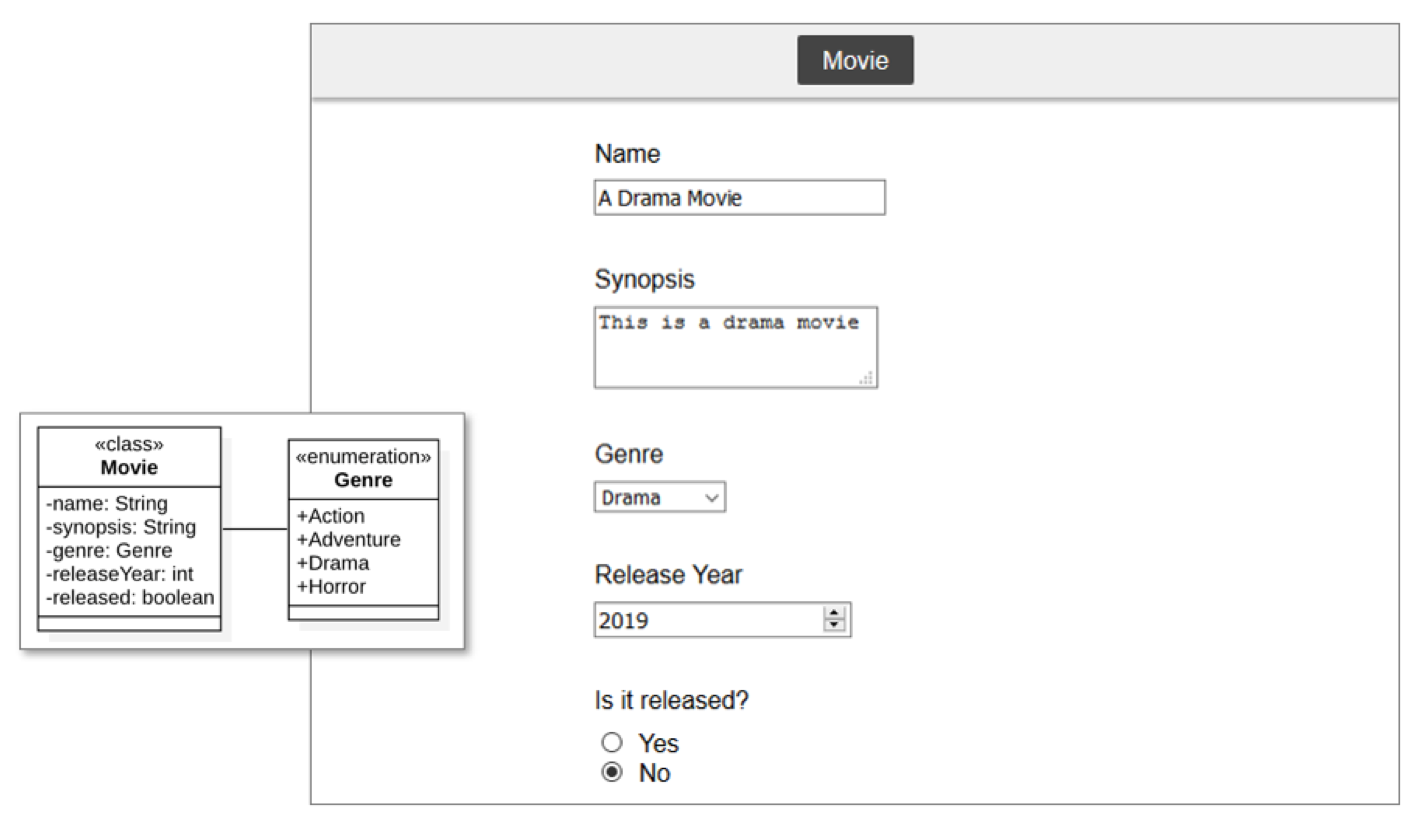}
\caption{Example of an interface generated by Pojo UI.}
\label{fig:interface}
\end{figure}

The second set of extensions provides support for the Pojo UI tool. This tool was developed together with this work to demonstrate the capacity the proposed architecture has to receive new extensions. It creates, through annotations of Java classes and their attributes, a basic graphical user interface, which displays panels that represent the classes, each presenting a form without functionality that has fields reflecting the attributes of the class and their respective types. The panels and fields can also have their labels, positions, and visibility modified through annotations. An example of the interface generated by the tool can be seen in Figure~\ref{fig:interface}. The extension for Pojo UI gives the graph vertices and edges that represent the panels and fields, in addition to their properties, which correspond to the classes and attributes extracted with the extension for Java language. To the transformation module, were added transformations that allow creating and removing panels and fields, as well as modifying the label, position, and visibility attributes of each one. Pojo UI is written in Java language and generates files in JavaScript, HTML, and CSS languages. The respective extension is developed using Java language for the extraction and injection modules and Prolog language for the transformation module. Implementations of the architecture and extensions can be found on GitHub\footnote{Github: Anonymous} 

\section{Evaluation}
\label{sec:evaluation}


To measure the effectiveness of Parthenos, metrics are collected from four scenarios that cover the functionality of Parthenos and are evaluated using precision, recall, and f-measure measures~\citep{Rijsbergen1979}.

\subsection{Evaluation Procedures}
\label{subsec:procedures}



\begin{figure}[h!]
\centering
\includegraphics[scale=0.19]{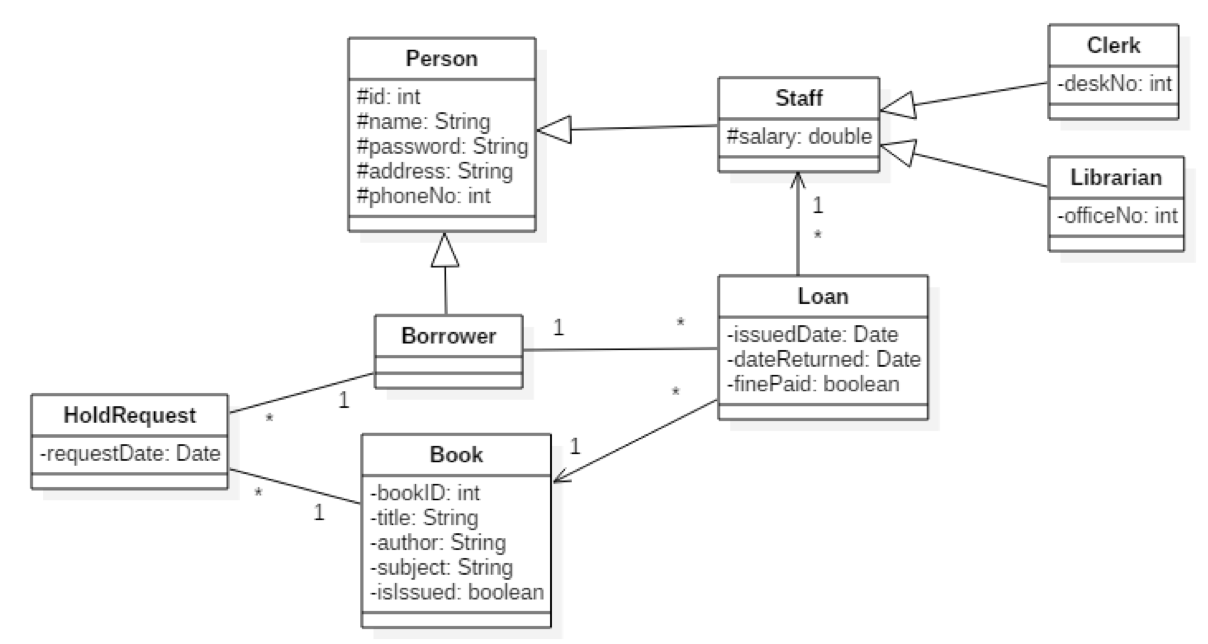}
\caption{Class diagram of the library management system.}
\label{fig:classes}
\end{figure}

As a way of evaluating Parthenos, four transformation scenarios were applied to a library management system available in a GitHub repository\footnote{Anonymous}. The system has a class diagram as seen in Figure~\ref{fig:classes} (adapted from $GitHub^{2}$) and has been adjusted to meet the scenarios\rq\space needs. Among the adjustments made, the system was structured to support Maven framework and the Pojo UI library was added as a dependency. In each performed scenario, the metrics obtained by applying a transformation to the model -- and, consequently, injecting the source code -- are compared to the metrics obtained from the model extracted from the source code after it is transformed manually, i.e., the scenario is applied to the source code as if a developer had done the change. This is so, because if a change is made manually by a human, and a new model is extracted thereof, that is, the source-of-truth we aim for when applying a respective automatic transformation.

\textbf{Evaluation scheme.} In this sense, some models are established, indicating the states of the system during the execution of a scenario. Such models are: 1) \textit{initial model} ($M_{A}$), which represents the state of the system before applying the scenario; 2) \textit{intermediate model} ($M_{B}$), which contains the elements, such as classes, attributes, facts, etc., generated by applying the transformations and injections related to the scenario; 3) \textit{source-of-truth (or desired) model} ($M_{AB}$), which represents the desired product obtained from the execution of this scenario; 4) \textit{obtained model} ($M_{C}$), which is the model obtained from applying the scenario. The greater the difference between $M_{C}$ and $M_{AB}$, the less the transformation was effective. If $M_{C}$ = $M_{AB}$, then the transformation reached its purpose.  We apply widely known software design metrics to characterize models, such as the number of classes, attributes, among others. The values obtained from the metrics are used to quantify the difference between the models --- thus being a way to evaluate the effectiveness of the proposed approach.


\textbf{Precision, recall, and f-measure.} The precision, recall, and f-measure measures are computed based on the mentioned metrics. These measures will be used as a way to calculate Parthenos\rq\space accuracy when executing each scenario. The precision measure indicates the accuracy of the obtained result. As can be seen in the formula $P = \frac{|M_{C} \cap M_{AB}|}{M_{C}}$, it calculates the percentage of additions made to the models that are relevant to the scenario being executed~\citep{chinchor1992}. The recall measure, in its turn, indicates whether all expected results were obtained, that is, what percentage of the expected additions were made to the models~\citep{chinchor1992}. The recall formula is \textit{R} = $\frac{|M_{C} \cap M_{AB}|}{M_{AB}}$.

Finally, the f-measure calculates the harmonic average between precision and recall measures, to combine them in a unified score that punctuates the effectiveness of Parthenos in the executed scenario. This score ranges from 0 to 1, with 1 being the best result and 0 the worst. The f-measure is important, as measuring the effectiveness of a scenario by looking only at precision or recall is incomplete and inaccurate, as they have a complementary role in this regard~\citep{chinchor1992}. The formula for f-measure is \textit{F} = 2.$\frac{P.R}{P+R}$.

\subsection{Metrics}
\label{subsec:metrics}


Table~\ref{tab:metrics} shows the metrics chosen to evaluate the accuracy of the transformations run by Parthenos. Syntax reckons the number of classes that compile without error. Semantics computes the number of classes with correct typing and functionality, i.e., classes which have not only the correct inheritance and field types, but also their functionalities reflect those of the desired scenario. KB is the number of elements presented by the abstract graph.



Note that both versions of the source code --- obtained ($M_{C}$) and desired ($M_{AB}$) --- and the abstract models are not directly compared in the following sections, as that would be inherently not feasible. Such comparisons are devised in this manner, due to the need to check not only the source code alone, but also the synchronicity between source code and the abstract model that represents it in each version. For this purpose, these metrics were created to be able to calculate quantitative performance values after the source code injection. The opacity that such quantitative evaluation can entail is mitigated by qualitative comments and analysis of each of the scenarios.

\begin{table}[!ht]
    \footnotesize
	\caption{List of the selected metrics}
	\label{tab:metrics}
	\centering
	\begin{tabular}{cl}
	\toprule
        \multicolumn{1}{c}{\textbf{Metric}} & 
        \multicolumn{1}{l}{\textbf{Description}}\\ 

    \midrule

    Classes         & Number of classes present in the system.    \\ 
    Attributes      & Number of attributes present in the system. \\ 

    Panels          & Number of panels processed in Pojo UI.      \\ 
    Fields          & Number of fields processed in Pojo UI.      \\ 

    Syntax          & Number of classes with correct syntax.      \\ 
    Semantics       & Number of classes with correct semantics.   \\ 

    KB              & Number of elements present in the abstract graph.    \\  

	\bottomrule
	\end{tabular}
\end{table}

\subsection{Evaluation Scenarios}
\label{subsec:evaluation_scenarios}


Before starting with the scenarios, the abstract graph model of the library management system was extracted, using the extraction module. The scenarios show the adaptive and perfective evolution of this system through applying transformations to the extracted model. The scenarios are described in the following sections and, for each, the obtained metrics are displayed.



\begin{figure}[ht]
\centering
\includegraphics[scale=0.18]{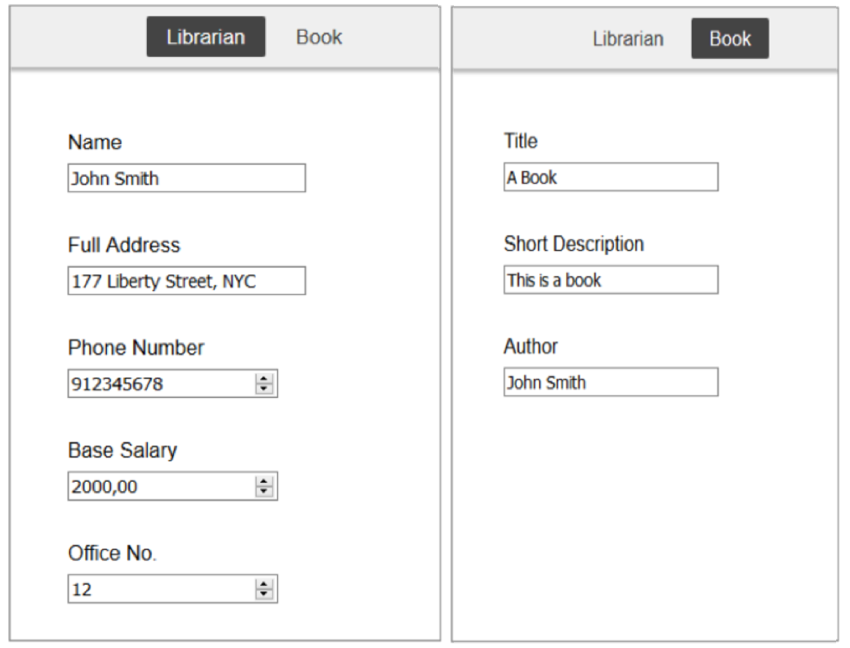}
\caption{Initial site generated for Pojo UI.}
\label{fig:site}
\end{figure}

\textbf{Scenario 1.} In this scenario, the system website is launched through Pojo UI. Figure~\ref{fig:site} shows the addition of panels for the classes \textit{Librarian} and \textit{Book}. Also, fields are added for the \textit{name, address, phoneNo, salary} and \textit{officeNo} attributes in the \textit{Librarian} class. Fields for \textit{title}, \textit{subject} and \textit{author} are added to the \textit{Book} class. Table~\ref{tab:cena1} presents the obtained results. In this experiment, four panels and eight fields were created according to the desired behavior. This first scenario did not have much business logic involved, but it aimed to assess how the code injection is able to create new UI from available entities.

\begin{table}[!ht]
    \footnotesize
	\caption{Metrics obtained in scenario 1.}
	\label{tab:cena1}
	\centering
	\begin{tabular}{lccccccc}
	\toprule
        \multicolumn{1}{c}{\textbf{Metrics}} & 
        \multicolumn{1}{c}{\textbf{$M_{A}$}} & 
        \multicolumn{1}{c}{\textbf{$M_{B}$}} & 
        \multicolumn{1}{c}{\textbf{$M_{AB}$}} &
        \multicolumn{1}{c}{\textbf{$M_{C}$}} &
        \multicolumn{1}{c}{\textbf{Precision}} &
        \multicolumn{1}{c}{\textbf{Recall}} &
        \multicolumn{1}{c}{\textbf{F-measure}}\\ 
    \midrule

Classes          & 8                    & 8                    & 8                    & 8                    & 1                   & 1               & 1                  \\ 
Attributes       & 17                   & 17                   & 17                   & 17                   & 1                   & 1               & 1                  \\ 

Panels           & 0                    & 4                    & 4                    & 4                    & 1                   & 1               & 1                  \\ 
Fields           & 0                    & 8                    & 8                    & 8                    & 1                   & 1               & 1                  \\ 

Syntax           & 8                    & 8                    & 8                    & 8                    & 1                   & 1               & 1                  \\ 
Semantics        & 8                    & 8                    & 8                    & 8                    & 1                   & 1               & 1                  \\ 

KB               & 1301                 & 68                   & 1369                 & 1369                 & 1                   & 1               & 1                  \\ 
                 & \multicolumn{1}{l}{} & \multicolumn{1}{l}{} & \multicolumn{1}{l}{} & \multicolumn{1}{l}{} & \multicolumn{2}{r}{\textit{Average:}} & 1                  \\ 
	\bottomrule
	\end{tabular}
\end{table}

\begin{figure}[h!]
\centering
\includegraphics[scale=0.35]{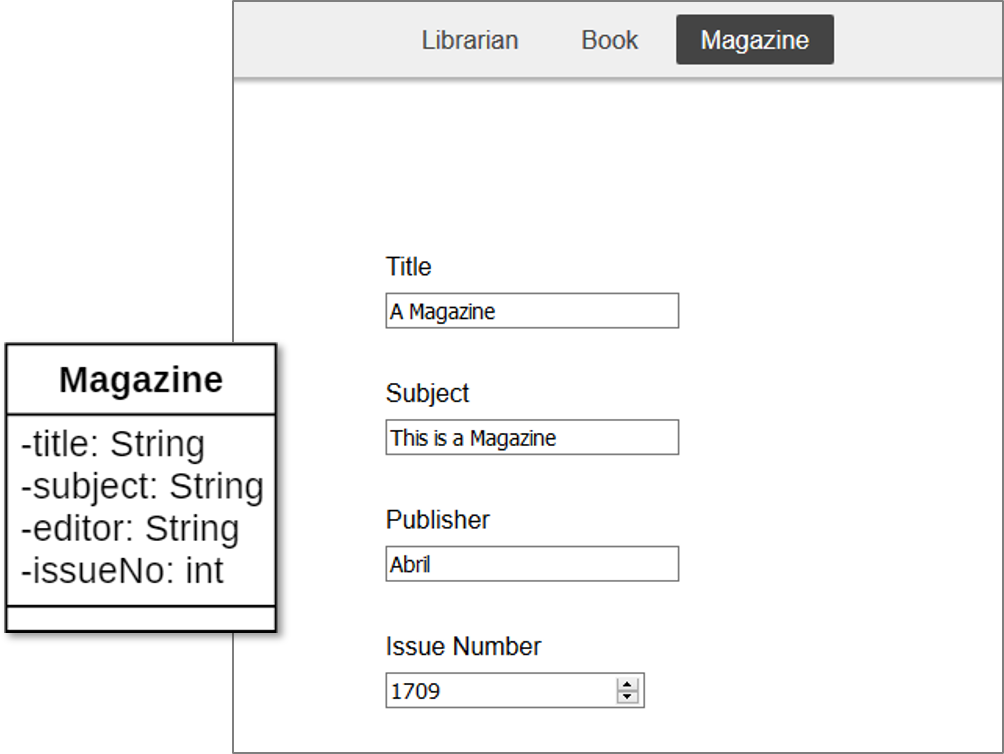}
\caption{Magazine class panel reflecting the generated class.}
\label{fig:scenario2}
\end{figure}

\textbf{Scenario 2.} The second scenario is a natural extension of the previous one. UI elements are created not for classes that are already present in the source code, but for a new one. The newly created class, its attributes and UI are added in runtime. In this scenario, the \textit{Magazine} class is added to the system. The~\textit{title}, \textit{subject}, \textit{publisher} and \textit{issueNo} attributes are added to the \textit{Magazine} class. A panel for the class and fields for the attributes are also created as seen in Figure~\ref{fig:scenario2}. Table \ref{tab:cena2} shows the collected results.

\begin{table}[!ht]
    \footnotesize
	\caption{Metrics obtained in scenario 2.}
	\label{tab:cena2}
	\centering
	\begin{tabular}{lccccccc}
	\toprule
        \multicolumn{1}{c}{\textbf{Metrics}} & 
        \multicolumn{1}{c}{\textbf{$M_{A}$}} & 
        \multicolumn{1}{c}{\textbf{$M_{B}$}} & 
        \multicolumn{1}{c}{\textbf{$M_{AB}$}} &
        \multicolumn{1}{c}{\textbf{$M_{C}$}} &
        \multicolumn{1}{c}{\textbf{Precision}} &
        \multicolumn{1}{c}{\textbf{Recall}} &
        \multicolumn{1}{c}{\textbf{F-measure}}\\ 
    \midrule
        
Classes          & 8                    & 1                    & 9                    & 9                    & 1                   & 1               & 1                  \\ 
Attributes       & 17                   & 4                    & 21                   & 21                   & 1                   & 1               & 1                  \\ 
Panels           & 4                    & 1                    & 5                    & 5                    & 1                   & 1               & 1                  \\ 
Fields           & 8                    & 4                    & 12                   & 12                   & 1                   & 1               & 1                  \\ 
Syntax           & 8                    & 1                    & 9                    & 9                    & 1                   & 1               & 1                  \\ 
Semantics        & 8                    & 1                    & 9                    & 9                    & 1                   & 1               & 1                  \\ 
KB               & 1369                 & 142                  & 1511                 & 1496                 & 1                   & 0.99            & 0.99               \\ 
                 & \multicolumn{1}{l}{} & \multicolumn{1}{l}{} & \multicolumn{1}{l}{} & \multicolumn{1}{l}{} & \multicolumn{2}{r}{\textit{Average:}} & 0.99              \\

	\bottomrule
	\end{tabular}
\end{table}

\begin{figure}[h!]
\centering
\includegraphics[scale=0.15]{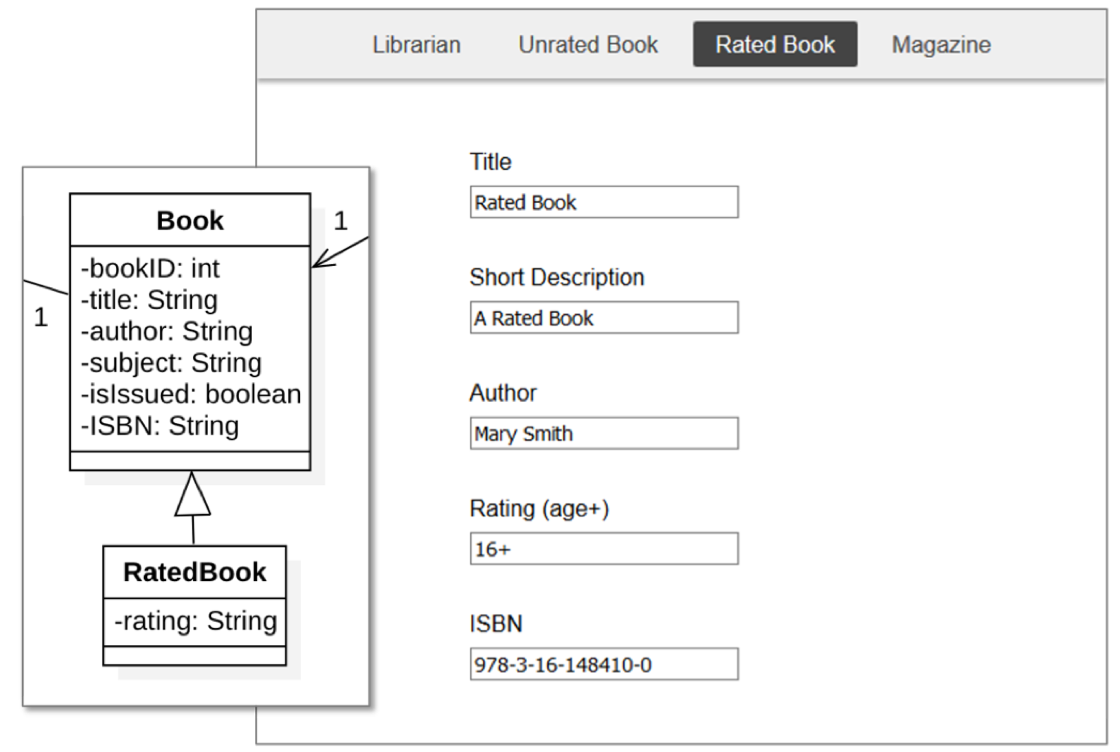}
\caption{Changes made for Scenario 3.}
\label{fig:scenario3}
\end{figure}

\textbf{Scenario 3.} This scenario creates a \textit{RatedBook} class that extends the \textit{Book} class by adding a \textit{rating} attribute. In addition, an \textit{ISBN} attribute is added to the \textit{Book} class. Looking at Figure~\ref{fig:scenario3}, it is possible to see that the respective panel and fields are added and cosmetic modifications are made: the \textit{Book} panel is renamed to ``Unrated Book'' and the \textit{RatedBook} panel is moved to before the \textit{Magazine} panel. Table \ref{tab:cena3} introduces the obtained results. The transformations introduced here involve two new aspects that were not explored previously, the first added class interacts with previous available classes through the mechanism of inheritance, and visual changes are applied to already existing UI elements.

\begin{table}[!ht]
    \footnotesize
	\caption{Metrics obtained in scenario 3.}
	\label{tab:cena3}
	\centering
	\begin{tabular}{lccccccc}
	\toprule
        \multicolumn{1}{c}{\textbf{Metrics}} & 
        \multicolumn{1}{c}{\textbf{$M_{A}$}} & 
        \multicolumn{1}{c}{\textbf{$M_{B}$}} & 
        \multicolumn{1}{c}{\textbf{$M_{AB}$}} &
        \multicolumn{1}{c}{\textbf{$M_{C}$}} &
        \multicolumn{1}{c}{\textbf{Precision}} &
        \multicolumn{1}{c}{\textbf{Recall}} &
        \multicolumn{1}{c}{\textbf{F-measure}}\\ 
    \midrule

Classes          & 9                    & 1                    & 10                   & 10                   & 1                   & 1               & 1                  \\ 
Attributes       & 21                   & 2                    & 23                   & 23                   & 1                   & 1               & 1                  \\ 

Panels           & 5                    & 1                    & 6                    & 6                    & 1                   & 1               & 1                  \\ 
Fields           & 12                   & 2                    & 14                   & 14                   & 1                   & 1               & 1                  \\ 

Syntax           & 9                    & 2                    & 10                   & 10                   & 1                   & 1               & 1                  \\ 
Semantics        & 9                    & 2                    & 10                   & 9                    & 1                   & 0.9             & 0.94               \\ 

KB               & 1496                 & 92                   & 1588                 & 1577                 & 1                   & 0.99            & 0.99               \\ 
                 & \multicolumn{1}{l}{} & \multicolumn{1}{l}{} & \multicolumn{1}{l}{} & \multicolumn{1}{l}{} & \multicolumn{2}{r}{\textit{Average:}} & 0.99               \\

	\bottomrule
	\end{tabular}
\end{table}



\begin{table}[!ht]
    \footnotesize
	\caption{Metrics obtained in scenario 4.}
	\label{tab:cena4}
	\centering
	\begin{tabular}{lccccccc}
	\toprule
        \multicolumn{1}{c}{\textbf{Metrics}} & 
        \multicolumn{1}{c}{\textbf{$M_{A}$}} & 
        \multicolumn{1}{c}{\textbf{$M_{B}$}} & 
        \multicolumn{1}{c}{\textbf{$M_{AB}$}} &
        \multicolumn{1}{c}{\textbf{$M_{C}$}} &
        \multicolumn{1}{c}{\textbf{Precision}} &
        \multicolumn{1}{c}{\textbf{Recall}} &
        \multicolumn{1}{c}{\textbf{F-measure}}\\ 
    \midrule

Classes          & 10                   & 0                    & 10                   & 10                   & 1                   & 1               & 1                  \\ 
Attributes       & 23                   & 0                    & 23                   & 23                   & 1                   & 1               & 1                  \\ 

Panels           & 6                    & 0                    & 6                    & 6                    & 1                   & 1               & 1                  \\ 
Fields           & 14                   & 0                    & 14                   & 14                   & 1                   & 1               & 1                  \\ 

Syntax           & 10                   & 0                    & 10                   & 10                   & 1                   & 1               & 1                  \\ 
Semantics        & 9                    & 0                    & 9                    & 9                    & 1                   & 1               & 1                  \\ 

KB               & 1577                 & 0                    & 1577                 & 1577                 & 1                   & 1               & 1                  \\ 
                 & \multicolumn{1}{l}{} & \multicolumn{1}{l}{} & \multicolumn{1}{l}{} & \multicolumn{1}{l}{} & \multicolumn{2}{r}{\textit{Average:}} & 1                  \\

	\bottomrule
	\end{tabular}
\end{table}

\textbf{Scenario 4.} It attempts to add a \textit{PuzzleBook} class that extends an \textit{UnratedBook} class (nonexistent, since it is actually called only \textit{Book}). After that, it tries to add a \textit{puzzleType} attribute to the \textit{PuzzleBook} class. The expected result in this scenario, differently from the other ones, is that both transformations fail and the models remain the same. Table~\ref{tab:cena4} presents the obtained results.


\subsection{Discussion and limitations}
\label{sec:discussion}


The \textit{f-measures} obtained high scores in all scenarios presented in Section~\ref{subsec:evaluation_scenarios}. Thus, this sounds that Parthenos was effective in extracting, transforming, and injecting the models, as well as covering scenarios with different needs in each one, avoiding syntactic and type-semantic errors. The extensible architecture allows systems written in different languages can be easily represented and transformed through an abstract graph model. This means that end-users can perform maintenance without programming knowledge. Table~\ref{tab:cena2} shows that not all the desired elements of the abstract graph model were generated. This is because the extraction module has a knowledge base processing step that adds metadata to the graph when completely new elements are added. This is the case with classes, causing problems for the semantics of graph types in specific scenarios. Fortunately, it did not hinder the execution of the rest of the scenarios.

We highlight that the precision, recall, and f-measure could be used to hide some specific problems, for instance. Suppose Parthenos excluded two classes, then created the same number of classes, but with totally different semantics. This result would be of nonpractical usage, however, the metrics would not capture this situation, indicating otherwise. The authors are aware of this limitation and a qualitative evaluation of the transformation was done to detect this type of problem.

\section{Conclusions and Future Work}
\label{sec:conclusion}

This work presented Parthenos, an implementation approach to MITRAS model that focuses on improving the source code injection step, ensuring syntactic and type-semantic correctness, in addition to ensuring synchronicity between the abstract graph model and the system's source code. The architecture proposed by Parthenos is robust and extensible, allowing systems written in different languages, using frameworks, libraries, and different tools to take advantage of the model proposed by MITRAS.



In future work, it is suggested to explore a way to inject source code by observing the semantics of languages as a whole. One example is implementing a module that semantically analyzes the injected files and provides the ability to recover from unambiguous semantic errors automatically. Another possibility is to provide the ability to generate code that adds new functionality (e.g., business logic) to systems through transformations. In both cases, the architecture and implementation made by Parthenos may serve as a basis for development.




\bibliographystyle{ACM-Reference-Format}
\bibliography{references}


\begin{thebibliography}{22}


\ifx \showCODEN    \undefined \def \showCODEN     #1{\unskip}     \fi
\ifx \showDOI      \undefined \def \showDOI       #1{#1}\fi
\ifx \showISBNx    \undefined \def \showISBNx     #1{\unskip}     \fi
\ifx \showISBNxiii \undefined \def \showISBNxiii  #1{\unskip}     \fi
\ifx \showISSN     \undefined \def \showISSN      #1{\unskip}     \fi
\ifx \showLCCN     \undefined \def \showLCCN      #1{\unskip}     \fi
\ifx \shownote     \undefined \def \shownote      #1{#1}          \fi
\ifx \showarticletitle \undefined \def \showarticletitle #1{#1}   \fi
\ifx \showURL      \undefined \def \showURL       {\relax}        \fi
\providecommand\bibfield[2]{#2}
\providecommand\bibinfo[2]{#2}
\providecommand\natexlab[1]{#1}
\providecommand\showeprint[2][]{arXiv:#2}

\bibitem[\protect\citeauthoryear{Antal, Havas, Siket, Besz{\'e}des, Ferenc, and
  Mihalicza}{Antal et~al\mbox{.}}{2016}]%
        {antal2016}
\bibfield{author}{\bibinfo{person}{G{\'a}bor Antal}, \bibinfo{person}{D{\'a}vid
  Havas}, \bibinfo{person}{Istv{\'a}n Siket}, \bibinfo{person}{{\'A}rp{\'a}d
  Besz{\'e}des}, \bibinfo{person}{Rudolf Ferenc}, {and}
  \bibinfo{person}{J{\'o}zsef Mihalicza}.} \bibinfo{year}{2016}\natexlab{}.
\newblock \showarticletitle{Transforming c++ 11 code to c++ 03 to support
  legacy compilation environments}. In \bibinfo{booktitle}{\emph{2016 IEEE 16th
  International Working Conference on Source Code Analysis and Manipulation
  (SCAM)}}. IEEE, \bibinfo{pages}{177--186}.
\newblock


\bibitem[\protect\citeauthoryear{Bischoff, Farias, Gon{\c{c}}ales, and
  Barbosa}{Bischoff et~al\mbox{.}}{2019}]%
        {bischoff2019integration}
\bibfield{author}{\bibinfo{person}{Vinicius Bischoff},
  \bibinfo{person}{Kleinner Farias}, \bibinfo{person}{Lucian~Jos{\'e}
  Gon{\c{c}}ales}, {and} \bibinfo{person}{Jorge Luis~Vict{\'o}ria Barbosa}.}
  \bibinfo{year}{2019}\natexlab{}.
\newblock \showarticletitle{Integration of feature models: A systematic mapping
  study}.
\newblock \bibinfo{journal}{\emph{Information and Software Technology}}
  \bibinfo{volume}{105} (\bibinfo{year}{2019}), \bibinfo{pages}{209--225}.
\newblock


\bibitem[\protect\citeauthoryear{Bondy, Murty, et~al\mbox{.}}{Bondy
  et~al\mbox{.}}{1976}]%
        {bondy1976}
\bibfield{author}{\bibinfo{person}{John~Adrian Bondy},
  \bibinfo{person}{Uppaluri Siva~Ramachandra Murty}, {et~al\mbox{.}}}
  \bibinfo{year}{1976}\natexlab{}.
\newblock \bibinfo{booktitle}{\emph{Graph theory with applications}}.
  Vol.~\bibinfo{volume}{290}.
\newblock \bibinfo{publisher}{Macmillan London}.
\newblock


\bibitem[\protect\citeauthoryear{Chagas, Farias, Gon{\c{c}}ales,
  Kupssinsk{\"u}, and Gluz}{Chagas et~al\mbox{.}}{2019}]%
        {chagas2019}
\bibfield{author}{\bibinfo{person}{Michael~William Chagas},
  \bibinfo{person}{Kleinner Farias}, \bibinfo{person}{Lucian Gon{\c{c}}ales},
  \bibinfo{person}{Lucas Kupssinsk{\"u}}, {and}
  \bibinfo{person}{Jo{\~a}o~Carlos Gluz}.} \bibinfo{year}{2019}\natexlab{}.
\newblock \showarticletitle{Hermes: A Natural Language Interface Model for
  Software Transformation}. In \bibinfo{booktitle}{\emph{Proceedings of the XV
  Brazilian Symposium on Information Systems}}. \bibinfo{pages}{1--8}.
\newblock


\bibitem[\protect\citeauthoryear{Chinchor and Sundheim}{Chinchor and
  Sundheim}{1993}]%
        {chinchor1992}
\bibfield{author}{\bibinfo{person}{Nancy Chinchor} {and}
  \bibinfo{person}{Beth~M Sundheim}.} \bibinfo{year}{1993}\natexlab{}.
\newblock \showarticletitle{MUC-5 evaluation metrics}. In
  \bibinfo{booktitle}{\emph{Fifth Message Understanding Conference (MUC-5):
  Proceedings of a Conference Held in Baltimore, Maryland, August 25-27,
  1993}}.
\newblock


\bibitem[\protect\citeauthoryear{de~Espindola, Majdenbaum, and
  Audy}{de~Espindola et~al\mbox{.}}{2004}]%
        {espindola2004}
\bibfield{author}{\bibinfo{person}{Rodrigo~Santos de Espindola},
  \bibinfo{person}{Azriel Majdenbaum}, {and} \bibinfo{person}{Jorge
  Luis~Nicolas Audy}.} \bibinfo{year}{2004}\natexlab{}.
\newblock \showarticletitle{Uma An{\'a}lise Cr{\'\i}tica dos Desafios para
  Engenharia de Requisitos em Manuten{\c{c}}{\~a}o de Software.}. In
  \bibinfo{booktitle}{\emph{WER}}. \bibinfo{pages}{226--238}.
\newblock


\bibitem[\protect\citeauthoryear{D’Avila, Farias, and Barbosa}{D’Avila
  et~al\mbox{.}}{2020}]%
        {d2020effects}
\bibfield{author}{\bibinfo{person}{Leandro~Ferreira D’Avila},
  \bibinfo{person}{Kleinner Farias}, {and} \bibinfo{person}{Jorge
  Luis~Vict{\'o}ria Barbosa}.} \bibinfo{year}{2020}\natexlab{}.
\newblock \showarticletitle{Effects of contextual information on maintenance
  effort: A controlled experiment}.
\newblock \bibinfo{journal}{\emph{Journal of Systems and Software}}
  \bibinfo{volume}{159} (\bibinfo{year}{2020}), \bibinfo{pages}{110443}.
\newblock


\bibitem[\protect\citeauthoryear{Farias, Garcia, and Lucena}{Farias
  et~al\mbox{.}}{2012}]%
        {farias2012evaluating}
\bibfield{author}{\bibinfo{person}{Kleinner Farias},
  \bibinfo{person}{Alessandro Garcia}, {and} \bibinfo{person}{Carlos Lucena}.}
  \bibinfo{year}{2012}\natexlab{}.
\newblock \showarticletitle{Evaluating the impact of aspects on inconsistency
  detection effort: a controlled experiment}. In
  \bibinfo{booktitle}{\emph{International Conference on Model Driven
  Engineering Languages and Systems}}. Springer, \bibinfo{pages}{219--234}.
\newblock


\bibitem[\protect\citeauthoryear{Farias, Garcia, and Lucena}{Farias
  et~al\mbox{.}}{2014}]%
        {farias2014effects}
\bibfield{author}{\bibinfo{person}{Kleinner Farias},
  \bibinfo{person}{Alessandro Garcia}, {and} \bibinfo{person}{Carlos Lucena}.}
  \bibinfo{year}{2014}\natexlab{}.
\newblock \showarticletitle{Effects of stability on model composition effort:
  an exploratory study}.
\newblock \bibinfo{journal}{\emph{Software \& Systems Modeling}}
  \bibinfo{volume}{13}, \bibinfo{number}{4} (\bibinfo{year}{2014}),
  \bibinfo{pages}{1473--1494}.
\newblock


\bibitem[\protect\citeauthoryear{Farias, Garcia, Whittle, and Lucena}{Farias
  et~al\mbox{.}}{2013}]%
        {farias2013analyzing}
\bibfield{author}{\bibinfo{person}{Kleinner Farias},
  \bibinfo{person}{Alessandro Garcia}, \bibinfo{person}{Jon Whittle}, {and}
  \bibinfo{person}{Carlos Lucena}.} \bibinfo{year}{2013}\natexlab{}.
\newblock \showarticletitle{Analyzing the effort of composing design models of
  large-scale software in industrial case studies}. In
  \bibinfo{booktitle}{\emph{International Conference on Model Driven
  Engineering Languages and Systems}}. Springer, \bibinfo{pages}{639--655}.
\newblock


\bibitem[\protect\citeauthoryear{Gersting}{Gersting}{2007}]%
        {gersting2006}
\bibfield{author}{\bibinfo{person}{Judith~L Gersting}.}
  \bibinfo{year}{2007}\natexlab{}.
\newblock \bibinfo{booktitle}{\emph{Mathematical structures for computer
  science}}.
\newblock \bibinfo{publisher}{Macmillan}.
\newblock


\bibitem[\protect\citeauthoryear{Gil and Orr{\`u}}{Gil and Orr{\`u}}{2017}]%
        {gil2017}
\bibfield{author}{\bibinfo{person}{Yossi Gil} {and} \bibinfo{person}{Matteo
  Orr{\`u}}.} \bibinfo{year}{2017}\natexlab{}.
\newblock \showarticletitle{The spartanizer: Massive automatic refactoring}. In
  \bibinfo{booktitle}{\emph{2017 IEEE 24th International Conference on Software
  Analysis, Evolution and Reengineering (SANER)}}. IEEE,
  \bibinfo{pages}{477--481}.
\newblock


\bibitem[\protect\citeauthoryear{Gon{\c{c}}ales, Farias, da~Silva, and
  Fessler}{Gon{\c{c}}ales et~al\mbox{.}}{2019}]%
        {gonccales2019measuring}
\bibfield{author}{\bibinfo{person}{Lucian Gon{\c{c}}ales},
  \bibinfo{person}{Kleinner Farias}, \bibinfo{person}{Bruno da Silva}, {and}
  \bibinfo{person}{Jonathan Fessler}.} \bibinfo{year}{2019}\natexlab{}.
\newblock \showarticletitle{Measuring the cognitive load of software
  developers: a systematic mapping study}. In \bibinfo{booktitle}{\emph{2019
  IEEE/ACM 27th International Conference on Program Comprehension (ICPC)}}.
  IEEE, \bibinfo{pages}{42--52}.
\newblock


\bibitem[\protect\citeauthoryear{Gon{\c{c}}ales, Farias, and
  da~Silva}{Gon{\c{c}}ales et~al\mbox{.}}{2021}]%
        {gonccales2021measuring}
\bibfield{author}{\bibinfo{person}{Lucian Gon{\c{c}}ales},
  \bibinfo{person}{Kleinner Farias}, {and} \bibinfo{person}{Bruno~C da Silva}.}
  \bibinfo{year}{2021}\natexlab{}.
\newblock \showarticletitle{Measuring the cognitive load of software
  developers: An extended Systematic Mapping Study}.
\newblock \bibinfo{journal}{\emph{Information and Software Technology}}
  (\bibinfo{year}{2021}), \bibinfo{pages}{106563}.
\newblock


\bibitem[\protect\citeauthoryear{Kahani, Bagherzadeh, Cordy, Dingel, and
  Varr{\'o}}{Kahani et~al\mbox{.}}{2019}]%
        {kahani2011}
\bibfield{author}{\bibinfo{person}{Nafiseh Kahani}, \bibinfo{person}{Mojtaba
  Bagherzadeh}, \bibinfo{person}{James~R Cordy}, \bibinfo{person}{Juergen
  Dingel}, {and} \bibinfo{person}{Daniel Varr{\'o}}.}
  \bibinfo{year}{2019}\natexlab{}.
\newblock \showarticletitle{Survey and classification of model transformation
  tools}.
\newblock \bibinfo{journal}{\emph{Software \& Systems Modeling}}
  \bibinfo{volume}{18}, \bibinfo{number}{4} (\bibinfo{year}{2019}),
  \bibinfo{pages}{2361--2397}.
\newblock


\bibitem[\protect\citeauthoryear{Knauer and Knauer}{Knauer and Knauer}{2019}]%
        {knauer2011}
\bibfield{author}{\bibinfo{person}{Ulrich Knauer} {and} \bibinfo{person}{Kolja
  Knauer}.} \bibinfo{year}{2019}\natexlab{}.
\newblock \bibinfo{booktitle}{\emph{Algebraic graph theory: morphisms, monoids
  and matrices}}. Vol.~\bibinfo{volume}{41}.
\newblock \bibinfo{publisher}{Walter de Gruyter GmbH \& Co KG}.
\newblock


\bibitem[\protect\citeauthoryear{Kupssinsk{\"u}}{Kupssinsk{\"u}}{2019}]%
        {kupssinski2019}
\bibfield{author}{\bibinfo{person}{Lucas~Silveira Kupssinsk{\"u}}.}
  \bibinfo{year}{2019}\natexlab{}.
\newblock \bibinfo{title}{MITRAS: modelo inteligente para
  transforma{\c{c}}{\~a}o de aplica{\c{c}}{\~o}es de software}.
\newblock
\newblock


\bibitem[\protect\citeauthoryear{Oliveira, Bischoff, Gon{\c{c}}ales, Farias,
  and Segalotto}{Oliveira et~al\mbox{.}}{2018}]%
        {oliveira2018}
\bibfield{author}{\bibinfo{person}{Anderson Oliveira},
  \bibinfo{person}{Vinicius Bischoff}, \bibinfo{person}{Lucian~Jos{\'e}
  Gon{\c{c}}ales}, \bibinfo{person}{Kleinner Farias}, {and}
  \bibinfo{person}{Matheus Segalotto}.} \bibinfo{year}{2018}\natexlab{}.
\newblock \showarticletitle{BRCode: An interpretive model-driven engineering
  approach for enterprise applications}.
\newblock \bibinfo{journal}{\emph{Computers in Industry}}  \bibinfo{volume}{96}
  (\bibinfo{year}{2018}), \bibinfo{pages}{86--97}.
\newblock


\bibitem[\protect\citeauthoryear{Rozenberg}{Rozenberg}{1997}]%
        {Rozenberg1997}
\bibfield{author}{\bibinfo{person}{Grzegorz Rozenberg}.}
  \bibinfo{year}{1997}\natexlab{}.
\newblock \bibinfo{booktitle}{\emph{Handbook of graph grammars and computing by
  graph transformation}}. Vol.~\bibinfo{volume}{1}.
\newblock \bibinfo{publisher}{World scientific}.
\newblock


\bibitem[\protect\citeauthoryear{Van~Rijsbergen}{Van~Rijsbergen}{2004}]%
        {Rijsbergen1979}
\bibfield{author}{\bibinfo{person}{Cornelis~Joost Van~Rijsbergen}.}
  \bibinfo{year}{2004}\natexlab{}.
\newblock \bibinfo{booktitle}{\emph{The geometry of information retrieval}}.
\newblock \bibinfo{publisher}{Cambridge University Press}.
\newblock


\bibitem[\protect\citeauthoryear{Vieira and Farias}{Vieira and Farias}{2020}]%
        {vieira2020usage}
\bibfield{author}{\bibinfo{person}{Roger~Denis Vieira} {and}
  \bibinfo{person}{Kleinner Farias}.} \bibinfo{year}{2020}\natexlab{}.
\newblock \showarticletitle{Usage of Psychophysiological Data as an Improvement
  in the Context of Software Engineering: A Systematic Mapping Study}. In
  \bibinfo{booktitle}{\emph{XVI Brazilian Symposium on Information Systems}}.
  \bibinfo{pages}{1--8}.
\newblock


\bibitem[\protect\citeauthoryear{Zafeiris, Poulias, Diamantidis, and
  Giakoumakis}{Zafeiris et~al\mbox{.}}{2017}]%
        {Zafeiris2017}
\bibfield{author}{\bibinfo{person}{Vassilis~E Zafeiris},
  \bibinfo{person}{Sotiris~H Poulias}, \bibinfo{person}{NA Diamantidis}, {and}
  \bibinfo{person}{Emmanouel~A Giakoumakis}.} \bibinfo{year}{2017}\natexlab{}.
\newblock \showarticletitle{Automated refactoring of super-class method
  invocations to the Template Method design pattern}.
\newblock \bibinfo{journal}{\emph{Information and Software Technology}}
  \bibinfo{volume}{82} (\bibinfo{year}{2017}), \bibinfo{pages}{19--35}.
\newblock


\end{thebibliography}





\end{document}